\documentclass[aps,twocolumn,showpacs]{revtex4}
\usepackage{graphicx}
\usepackage{amsmath}

\begin{document}

\title{Superlattice formed by quantum-dot sheets: density of states and IR absorption}
\author{F. T. Vasko}
\email{fedirvas@buffalo.edu}
\author{V. V. Mitin}
\affiliation{Department of Electrical Engineering, University at Buffalo, Buffalo, NY 14260-1920, USA}
\date{\today}

\begin{abstract}
Low-energy continuous states of electron in heterosrtucture with periodically placed quantum-dot sheets are studied theoretically. The Green's function of electron is governed by the Dyson equation with the self-energy function which is determined the boundary conditions at quantum-dot sheets with weak damping in low-energy region. The parameters of superlattice formed by quantum-dot sheets are determined using of the short-range model of quantum dot. The density of states and spectral dependencies of the anisotropic absorption coefficient under mid-IR transitions from doped quantum dots into miniband states of superlattice strongly depend on dot concentration and on period of sheets. These dependencies can be used for characterization of the multi-layer structure and they determine parameters of different optoelectronic devices exploiting vertical transport of carriers through quantum-dot sheets. 
\end{abstract}
\maketitle

\section{Introduction}
Heterostructures formed by quantum dot (QD) sheets are widely investigated and used in different devices, such as lasers, photodetectors, and solar cells, see \cite{1,2,3} for review. In such heterostructures, not only the additional localized states of electrons captured into QDs should be taken into account but also the continuous electronic states, which are subjected to reflections on periodically placed QD sheets, should be modified significantly. Such a periodical perturbation gives rise to a superlattice (SL) with energy spectrum formed by gaps between allowed minibands. In contrast to the standard case, \cite{3a} an additional damping of electronic states takes place due to scattering on inhomogeneties of QD sheets stemming from a random in-plane distribution of QDs. But such a damping appears to be weak for low-energy region. As a result, SL effect should be essential near the edge of interband absorption in host material, which is proportional to the density of states of SL, or under IR transitions from doped QDs into miniband states. To the best of our knowledge, these phenomena were not considered based on a simultaneous description of SL minibands and damping effects in spite of the structures under consideration are routinely used in different optoelectronic devices. At the same time the opposite case of 3D ordering of the closely spaced QDs, when SL is formed as a result of the tunneling mix between intra-QD states, was analyzed \cite{4} and demonstrated experimentally, see \cite{5,6} and references therein. Because of this, it is important and timely to develop an adequate theory of low-energy electrons interacting with the periodically-placed QD sheets and to study the optical response of SL which can be used for characterization of structures under consideration and for description of different optoelectronic devices.

In this paper we study low-energy electronic states, with energies in the vicinity of the conduction band extremum, in heterosrtuctures formed by QD sheets of period $l$ using the effective-mass equations for the Green's function averaged over randomly placed QDs in each sheet. In contrast to the standard theoretical description based on the averaging over 3D or 2D space, \cite{7}, here we perform the averaging over QD sheets with the identical statistical characteristics. As a result, we obtain the inhomogeneous along SL axis Dyson equation where the self-energy function can be replaced by the boundary conditions at QD sheets. Since the damping of the low-energy states is weak, one can consider SL which characteristics are determined by an effective potential localized at the sheet positions, $z=nl$, $n=0,\pm 1,\ldots$. The strength of this potential is determined by the concentration of QDs and shape of QD potential. With respect to low-energy states, QD can be considered as a short-range defect (which are widely investigated during past 50 years, see Refs. 9) if the low-energy interval under consideration is smaller than the QD binding energy. 

The density of states in SL depends on the period $l$ and on the parameter determined by a strength of QD's potential described within the short-range approximation. Spectral dependencies of interband absorption between the heavy-hole and SL states are proportional to the density of states in $c$-band. In addition, the anisotropic absorption coefficient, originated due to mid-IR transitions from the doped QD ground state into the miniband states of SL, is obtained through the QD concentration and the SL parameters. We found that the efficiency of mid-IR photoexcitation is comparable to the contribution of wetting layers formed under QD sheets \cite{9} if doping levels are the same. But the spectral dependencies are very different for these two mechanisms. Thus, it is demonstrated that the results obtained can be used for characterization of structure under consideration. It is more important that the SL parameters determine a mechanism of vertical transport for underbarrier electrons which is a key process in different optoelectronic devices exploiting multi-QD sheets. Similar mechanism of transport through underbarrier states of IR photodetectors formed by GaAs/AlGaAs-based SL was considered in Ref. 11. 

The paper is organized as follows. In Sect. II we describe the model of periodical sheets formed by randomly placed QDs and evaluate the Green's function averaged over random positions of QDs. SL effects on the density of states and on the process of anisotropic photoexcitation of QDs are considered in Sect. III. List of assumptions used and concluding remarks are presented in the last section. Appendix contains the justification of the effective SL approach employed in the calculations performed.

\section{Model}
The electronic states near $c$-band extremum of heterostructure, which is formed by QD sheets placed in host material, are described by the effective mass Hamiltonian
%1
\begin{equation}
\hat H =\frac{\hat p^2}{2m} + \sum\limits_{rk} {u\left( {{\bf r} - {\bf R}_{rk} } \right)} , 
\end{equation}
where $\hat{\bf p}$ is the 3D momentum operator, $m$ is the effective mass, and $u({\bf r}-{\bf R}_{rk})$ is the potential energy of QD placed at coordinates ${\bf R}_{rk} =({\bf x}_{rk},rl)$. Here $r$ labels sheet ($r=0,\pm 1,\pm 2,\ldots$) placed with the period $l$ and $k$ stands for position of QD over the $r$th sheet given by 2D random coordinate ${\bf x}_{rk}$ ($k=1,2\ldots N$ where $N$ is number of QDs over each sheet with the normalization area $L^2$). Electron of energy $E$ is described by the Green's function ${\cal G}_E \left({\bf r},{\bf r}'\right)$ governed by the equation
%2
\begin{equation}
\left( {E + i\lambda  - \hat H} \right){\cal G}_E \left({\bf r},{\bf r}'\right) = \delta \left( {{\bf r} - {\bf r}'} \right)
\end{equation}
with $\lambda\to +0$ and the 3D $\delta$-function $\delta (\Delta{\bf r})$. Below we consider the averaged over all QD positions Green's function $G_E \left({\bf r},{\bf r}'\right)=\left\langle{\cal G}_E \left({\bf r},{\bf r}'\right)\right\rangle$ where the averaging over $r$th sheet is performed according to \cite{7}
%3
\begin{equation}
\left\langle  \cdots  \right\rangle _r  = \frac{1}{L^{2N}}\int {d{\bf x}_{r1}  \cdots } \int {d{\bf x}_{rN}  \cdots } 
\end{equation}
and $\left\langle\ldots\right\rangle$ includes the averaging over all sheets.

Using the $({\bf p},z)$-representation ($\bf p$ is 2D momentum) one obtains the Dyson equation governing the averaged Green's function as follows
%4
\begin{eqnarray}
G_{Ep} \left( z,z' \right) = g_{Ep} \left( z-z' \right) ~~~~~  \\
+ \int {dz_1} \int {dz_2 g_{Ep} \left( z-z_1 \right)}\Sigma _{Ep} \left( {z_1 ,z_2 } \right)G_{Ep} 
\left( z_2 ,z' \right) .   \nonumber
\end{eqnarray} 
Here $g_{Ep} \left( z-z' \right)$ is the free Green's function which is governed by Eq. (2) with the Hamiltonian $\hat{p}^2/2m$, so that
%5
\begin{equation}
g_{Ep} (\Delta z) = \frac{1}{\hbar }\sqrt {\frac{m}{{2(\varepsilon _p  - E)}}} \exp \left( -\frac{ \sqrt {2m(\varepsilon _p  - E)} \Delta z}{\hbar } \right) ,
\end{equation}
if $\varepsilon_p >E$ and the imaginary factor $i\sqrt {E-\varepsilon _p}$ should be used in (5) if $\varepsilon_p <E$. Within the self-consistent Born approximation, the self-energy function $\Sigma _{Ep} \left( {z_1 ,z_2 } \right)$ in Eq.(4) is given by
%6
\begin{eqnarray}
\Sigma _{Ep} \left( {z_1 ,z_2 } \right) \simeq \frac{n_{QD}}{{L^2 }}\sum\limits_{r{\bf p}_1 } u\left(\frac{{\bf p} - {\bf p}_1}{\hbar},z_1  - rl\right) \\
\times G_{Ep_1 } \left( {z_1 ,z_2 } \right) u\left(\frac{{\bf p}_1 - {\bf p}}{\hbar},z_2  - rl\right) +\ldots ,
 \nonumber
\end{eqnarray} 
where $u({\bf q},z)$ is the 2D Fourier transform of  $u({\bf r})$ and $n_{QD}$ is the QD concentration over sheet which does not dependent on $r$, i.e. we consider identical QD sheets.

Further, we restrict ourselves by the low-energy region where scattering on a QD can be described by the short-range potential $u( {\bf r})\approx U\Delta ({\bf r})$ with the form-factor $\Delta ({\bf r})$ localized in volume $\sim a^3$ ($a$ stands for the characteristic size of QD). We also neglect high-order corrections to the self-energy function (6), see below the diagram expansion of Fig. 4 and discussion in Appendix. Since the kernel (6) is located near QD sheets with $z_{1,2}\sim rl$ and the Green's functions vary over scales $\hbar /\sqrt{2m|E-\varepsilon_p |}$, the integral equation (4) is transformed into the finite-difference one:
%7
\begin{eqnarray}
G_{Ep} (z,z') =g_{Ep} (z - z')  \\
+ \Lambda _{Ep} \sum\limits_r {g_{Ep} (z- rl)G_{Ep} (rl,z')}  .  \nonumber
\end{eqnarray} 
The self-energy function (6) is written here through the factor
%8
\begin{equation}
\Lambda_{Ep}  = \frac{{n_{QD} }}{{L^2 }}\sum\limits_{{\bf p}_1 } {G_{Ep_1 } (rl,rl)} \left| {\int {d\Delta zu\left( {\frac{{\bf p}-{\bf p}_1}{\hbar },\Delta z} \right)} } \right|^2 ,
\end{equation}
which is the same for any QD sheet [we moved $\Sigma_r\ldots$ from Eq. (6) to Eq. (7)]. Instead of Eq. (7), one can determine $G_{Ep} (z,z')$ from Eq. (2) with the free Hamiltonian $\hat{p}^2/2m$ and describe the QD sheet effect adding the boundary conditions
%9
\begin{eqnarray}
\frac{{\hbar ^2 }}{2m}\left[\frac{d}{{dz}}G_{Ep} (z,z')\right]_{z=rl-0}^{z=rl+0} = \Lambda_{Ep} G_{Ep} (rl,z'), \\  G_{Ep} (z,z')\left|_{z=rl-0}^{z=rl+0}  \right. = 0  ~~~~~~~~~ \nonumber
\end{eqnarray}
at sheet positions $z=rl$. This result was evaluated after acting of the operator 
$E+i\lambda -\hat p^2/2m$ on the integral Dyson equation (4) and subsequent integration of the intergo-differential equation obtained over the QD positions $(rl-0,rl+0)$.

Within the second-order Born approximation we use $G_{Ep} \left( rl ,rl \right)\simeq g_{Ep} (0)$ in the self-consistent equation (8), see Ref. 8 for detais, and the momentum-independent factor $\Lambda_E$ in Eq. (9) takes the form
%10
\begin{equation}
\Lambda_E =\Lambda\left( 1+i\sqrt{\frac{E}{\varepsilon_a}}\right)  ,
~~~~ \Lambda\equiv\frac{{n_{QD} }}{2}U^2\overline{\rho}_{\varepsilon_a} .
\end{equation}
Here we estimate $\Lambda$  for the case of short-range defect within the Koster-Slater approach \cite{10} and $\overline{\rho}_E$ is the 3D density of states  which is taken at the cut-off energy  $\varepsilon_a \sim (\pi \hbar /a)^2 /2m$. Since $E\ll \varepsilon_a$, damping of low-energy states is weak and one can replace the complex boundary condition (9) by the effective potential energy $-\Lambda\sum_r\delta_a (z-rl)$ with $\delta_a (\Delta z)$ localized in the interval $|\Delta z|<a$, so that in the framework of the effective SL approach $G_{Ep} (z,z')$ is governed by the one-dimensional equation:
%11
\begin{eqnarray}
\left( E+i\lambda -\varepsilon_p -\hat{H}_\bot\right) G_{Ep}(z,z')=\delta (z-z') , \\
\hat{H}_\bot =\frac{\hat p_z^2}{2m}-\Lambda\sum\limits_r \delta_a\left( z-rl\right)  \nonumber
\end{eqnarray}
with the electron effective mass in the GaAs matrix, $m$. Thus, the Green's function is expressed using the standard relation \cite{7} between $G_{Ep}(z,z')$ and the solutions of the eigenstate problem for SL, \cite{11} $\hat{H}_\bot\psi_{z}^{(np_\bot )}=\varepsilon_{np_\bot}\psi_{z}^{(np_\bot )}$. The last equation determines the dispersion relations $\varepsilon_{np_\bot}$ and the eigenfunctions $\psi_{z}^{(np_\bot )}$. Here $p_\bot$ is quasimomentum ($|p_\bot |<\pi\hbar /l$), $n$ labels minibands, and the wavefunction takes form
%12
\begin{equation}
\psi _z^{(np_\bot )} =\psi_{np_\bot}\left( {e^{ik_{np_\bot}z} -R_{np_\bot} 
e^{-ik_{np_\bot}z} } \right) ,
\end{equation}
where the reflection coefficient and the normalization factor, $R_{np_\bot }$ and $\psi_{np_\bot}$, are expressed through $p_\bot$ and $k_{np_\bot}$. \cite{12} The energy $\varepsilon_{np_\bot}=(\hbar k_{np_\bot})^2/2m$ is founded from the dispersion equation
%13
\begin{equation}
\cos \frac{p_\bot l}{\hbar} = \cos k_{np_\bot}l -\frac{K}{k_{np_\bot}}
\sin k_{np_\bot} l , 
\end{equation}
which is written through the characteristic wave vector, $K=\Lambda m/\hbar^2\sim\pi^3 n_{QD}a/2$, see Ref. 13 for details.
%f1
\begin{figure}[tbp]
\begin{center}
\includegraphics[scale=1.1]{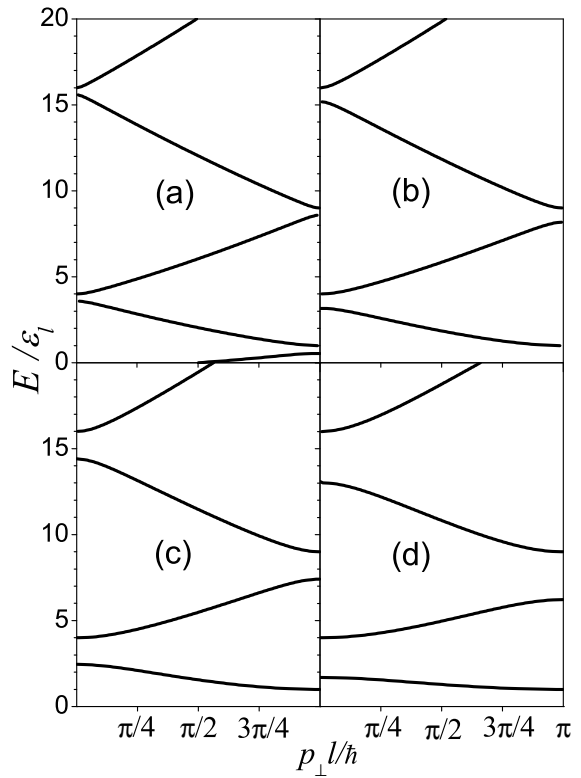}
\end{center}
\addvspace{-0.5 cm}
\caption{Miniband energy spectra, $E/\varepsilon_l$ versus $p_\bot  l/\hbar$, of the effective SL determined by Eq. (13) for $Kl=$1 (a), 2 (b), 4 (c), and 8 (d). } 
\end{figure}

The dispersion relations for lower minibands determined by Eq. (13) are shown in Fig. 1 for dimensionless parameter $Kl$ varied between 1 and 8 when the transformation from the weakly-coupled SL (if $Kl\leq$2) to the tight-binding regime of coupling (if $Kl>$4) takes place. The characteristic energy $\varepsilon_l =(\pi\hbar /l)^2/2m$ is about 3.2 meV for SL of period $l=$40 nm. For SL formed by InAs QDs embedded by GaAs matrix $Kl\approx$3.1 if $n_{QD}\simeq 5\times 10^{10}$ cm$^{-2}$. As a result, minigaps exceed 5 meV for the tight-binding regime, see Figs. 1c and 1d when dispersion laws are close to cosine and sine dependencies, for odd and even $n$ respectively. For the weakly-coupled SLs the dispersion laws are formed by parabolic curves modified near $p_\bot l/\hbar =0, \pi$ with gaps $\sim$1 meV, see Figs. 1a and 1b. In contrast to SL corresponding to under-barrier tunneling regime, \cite{11} if $Kl\leq$1.5 one obtains the lowest miniband at finite $p_\bot l/\hbar$ only, as it is shown in Fig. 1a. This is because of absence of solution for Eq. (13) at $p_\bot\to 0$ and $k_{np_\bot}l\ll 1$. Such a peculiarity change the density of states and the edge of mid-IR absorption if $Kl\leq 1.5$, see Figs. 2a and 3a below.

\section{Results}
Using the model described above, we consider in this section the density of states in SL formed by QD sheets, and calculate the absorption coefficient under mid-IR photoexcitation from ground levels of doped QDs into miniband states of SL.

\subsection{Density of states}
The density of states is introduced through the averaged Green's function by the standard formula \cite{7}
%14
\begin{eqnarray}
\rho_E =-\frac{2}{\pi L^3}{\rm Im}\int d{\bf r} \left\langle{\cal G}_E \left( {\bf r},{\bf r} \right) \right\rangle \\
\simeq \frac{2}{L^3} \sum\limits_{np_\bot{\bf p}}\delta (E-\varepsilon_p -\varepsilon_{np_\bot}) ,
\nonumber
\end{eqnarray}
where 2 is due to spin degeneracy and $L^3$ is the normalization volume. The lower expression is obtained for the case of negligible damping in Eq. (10) using of the effective SL approach determined by Eqs. (11)-(13), see the energy spectra plotted in Fig. 1. The integration of $\delta$-function over $\bf p$ gives the 2D density of states, $\rho_{2D}$, and after integration of $\theta$-function over $p_\bot$ the density of states should be replaced by constant if $E$ belongs to $\bar n$th gap. In $\bar n$th miniband (below $\bar n$th gap), the integral over $p_\bot$ should be taken over the interval $(0,p_E)$ where $p_E$ is found as a root of the equation $E=\varepsilon_{\bar np_E}$. As a result, $\rho_E$ takes the form:
%15
\begin{equation}
\rho_E =\frac{\rho_{2D}}{l}\left\{ \begin{array}{*{20}c}
\bar n, & E \subset  \bar n{\rm th ~ gap}  \\
\bar n-1+p_E l/(\pi\hbar ) , & E \subset  \bar n{\rm th ~band} 
\end{array} \right.
\end{equation}
and a shape of $\rho_E$ is determined by the gap-induced steps with transitions between them determined by the miniband dispersion laws.
%f2
\begin{figure}
%[tbp]
\begin{center}
\includegraphics[scale=1.2]{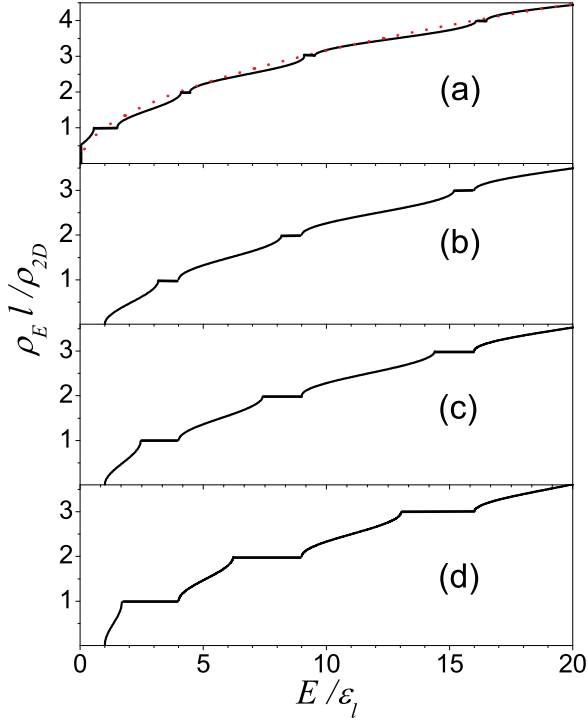}
\end{center}
\addvspace{-0.5 cm}
\caption{ Normalized density of states $\rho_E l/\rho_{2D}$ versus $E/\varepsilon_l$ given by Eq. (15) for the parameters used in panels (a-d) of Fig. 1. Dotted curve in upper panel corresponds to the 3D density of states $\propto\sqrt{E}$ if SL effect is negligible, $\Lambda\to 0$.}
\end{figure}

In Fig. 2 we plot the dimensionless density of states, in units $\rho_{2D}/l$, for the same parameters as in Fig. 1. For the weak coupling regime, the jump of $\rho_E$ at $E\to 0$ appears due to the cut-off of the lowest miniband at finite $p_\bot l/\hbar$, c. f. Figs. 1a and 2a at $E/\varepsilon_l\leq 2$. With increasing of $Kl$ under transition to the tight-binding regime, the energy-independent gap contributions to the density of states increase and $\rho_E$ between these steps is transformed from $\propto\sqrt{E}$ dependency shown by dotted curve in Fig. 2a to the arccosine dependencies. In addition, the bottom of lowest subband is shifted to energies $\sim\varepsilon_l$. Since $\rho_E$ is connected directly to the shape of interband optical spectra, see Ref. 13, the step-like dependencies permit one to extract $Kl$ value which determine the bandstructure of SL according to Eq. (13). 

Let us compare the energy scale of SL effect, determined by $\varepsilon_l$, and the SL effect due to the wetting layer contribution analyzed in Refs. 10. For the parameters given at the end of Sect. II, one obtains that the contribution of QD sheet with $n_{QD}=5\times 10^{10}$ cm$^{-2}$ is reduced $\sim$2 times in comparison with the wetting layer effect if levels of electron doping are the same. Thus, an interplay of both mechanisms should take place for $n_{QD}\geq 10^{11}$ cm$^{-2}$. For such a case, the interband optical spectra should be dependent on both the QDs contributions and the wetting layer contributions.
 
\subsection{Photoionization}
The anisotropic absorption coefficients $\alpha^{||}_\omega$ and $\alpha^{\bot}_\omega$ are determined from the general Kubo formula as follows:
%16
\begin{eqnarray}
\alpha^{||,\bot}_\omega =\frac{8(\pi e)^2}{\sqrt\epsilon c\omega L^3}\sum\limits_{\delta \delta '}\left[ f(\varepsilon_\delta )-f(\varepsilon_\delta +\hbar\omega )\right]  
\nonumber \\
\times\left| (\delta |{\bf e}_{||,\bot}\cdot\hat{\bf v}|\delta ')\right|^2\delta \left(\varepsilon_\delta -\varepsilon_{\delta '}+\hbar\omega\right) ,
\end{eqnarray}
where $\epsilon$ is the dielectric permittivity of the host semiconductor and the matrix element $\left| (\delta |{\bf e}_{||,\bot}\cdot\hat{\bf v}|\delta ')\right|^2$ corresponds to transitions between $\delta$- and $\delta '$-states of energies $\varepsilon_\delta$ and $\varepsilon_{\delta '}$ under radiation with polarization orts ${\bf e}_{||,\bot}$. We use the equilibrium distributions $f(\varepsilon_\delta )\to 1$ and $f(\varepsilon_\delta +\hbar\omega )\to 0$ because the only localized states are populated at temperatures lower the binding energy $|E_0|$. Due to the in-plane isotropy of the problem, we separate the cases of $s$- and $p$-polarized radiation corresponding to the polarization orts ${\bf e}_\|$ and ${\bf e}_z$. Neglecting the overlap between QD states and taking the ground state wave functions $\Psi_P$ in the momentum representation ($\bf P$ is 3D momentum) we transform Eq. (16) into:
%17
\begin{eqnarray}
\left| {\begin{array}{*{20}c}
\alpha_\omega^{\|} \\  \alpha_\omega^\bot \\
\end{array}} \right| =-\frac{4\pi e^2}{\sqrt\epsilon c\omega m^2L^9}\sum\limits_{{\bf PP}'}\Psi_P\Psi_{P'}^* \int d{\bf r}\int d{\bf r} ' \nonumber \\
\times e^{i({\bf Pr} - {\bf P}'{\bf r}')/\hbar }\left|\begin{array}{*{20}c}
({\bf e}_{\|} {\bf P})({\bf e}_{\|} {\bf P}')  \\
{\hat p}_z{\hat p}_{z'}^+  \\
\end{array}\right| K_{\Delta{\bf p},E_0 +\hbar\omega}({\bf r}',{\bf r}) . ~~~ 
\end{eqnarray}
The contribution of miniband states are described here through the average of the exact Green's function ${\cal G}_E ({\bf r}',{\bf r})$ with the exponential factor corresponding to random QD positions (here $\Delta {\bf p}\equiv {\bf P}-{\bf P'}$):
%18
\begin{equation}
K_{\Delta {\bf p},E} ({\bf r}',{\bf r}) = \left\langle {\sum\limits_{rk} {e^{i\Delta {\bf pR}_{rk} /\hbar }{\rm Im}{\cal G}_E ({\bf r}',{\bf r})} } \right\rangle 
\end{equation}
which is analyzed in the Appendix. Within the low-order approach, the correlation function (18) takes the form:
%19
\begin{equation}
K_{\Delta {\bf p},E} ({\bf r}',{\bf r}) \approx N_{QD} \frac{L}{l}\delta _{\Delta {\bf p},0} {\rm Im} G_E ({\bf r}',{\bf r})  ,
\end{equation}
where $N_{QD}L/l$ is the total number of QDs in the normalization volume $L^3$ and the averaged Green's function $G_E ({\bf r}',{\bf r})$ was considered in Sect. II.
%f3
\begin{figure}[tbp]
\begin{center}
\includegraphics{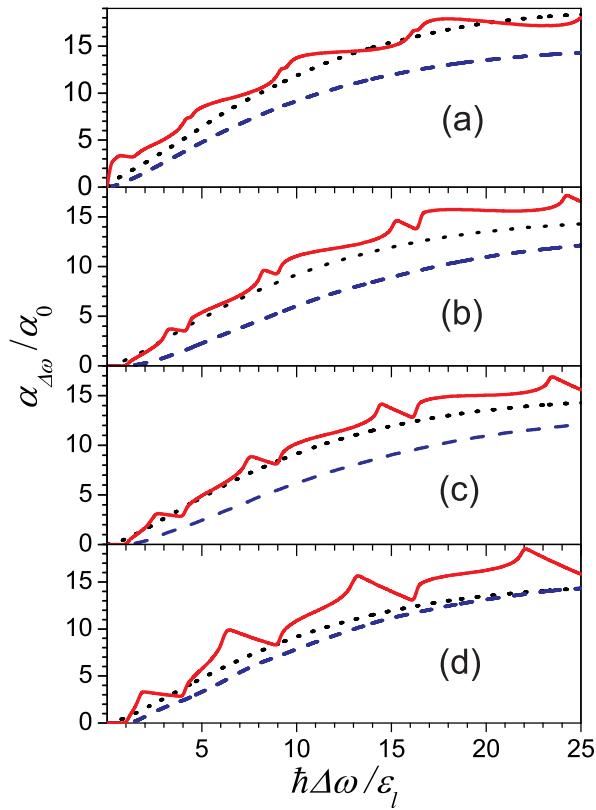}
\end{center}
\addvspace{-0.5 cm}
\caption{(Color online) Spectral dependencies of dimensionless absorption coefficients determined by Eqs. (20) and (21) for the same conditions as in Figs. 1 and 2. Solid and dashed curves correspond to the perpendicular and parallel polarizations, respectively. Dotted curves correspond to the case $\Lambda\to 0$, when SL effect is negligible.}
\end{figure}

Using the ground state wave function $\Psi_P$ written in the Koster-Slater approach \cite{10} and neglecting the damping correction in Eq. (10) we transform Eq. (17) as follows
%20
\begin{eqnarray}
\left| {\begin{array}{*{20}c}
   {\alpha_\omega^{||} }  \\   {\alpha _\omega ^ \bot  }  \\
\end{array}} \right| = \frac{{(2\pi e)^2 n_{QD} }}{{\sqrt \epsilon  c\omega m^2 lL^3 }}\sum\limits_{{\bf p}p_ \bot  } {\left| \Psi_P \right|}^2 \left| {\begin{array}{*{20}c}
   p^2/2  \\    {p_ \bot ^2 }  \end{array}} \right| ~~~ \\
\times\sum_{n\bar p_\bot}\left| {\frac{2}{l}\int\limits_{ - l}^l {dze^{ - ip_ \bot  z/\hbar } \psi _z^{(n\bar p_ \bot  )} } } \right|^2 \delta \left( {\hbar \Delta \omega  - \varepsilon _{n\bar p_\bot  }  - \varepsilon _p } \right) . \nonumber
\end{eqnarray}
Here ${\bf P}\equiv ({\bf p},p_{\bot})$ and we have replaced $G_{Ep}(z',z)$ from Eq. (19) using the wave function (12). In the expressions for $\alpha_\omega^{||,\bot}$ integrals over $\bf p$-plane and over $z$ are taken analytically and the spectral dependencies of IR absorption are obtained after the double numerical integrations over the transverse momenta $p_\bot$ and $\bar p_\bot$. The dimensionless spectral dependencies are plotted in Fig. 3 for the same conditions as in Figs. 1 and 2. The characteristic absorption $\alpha_0$ is given by
%21
\begin{equation}
\alpha_0 =\frac{(4e)^2 n_{QD}}{c\sqrt{\epsilon m|E_0 |/2}}\left(\frac{\varepsilon_l} {E_0} \right)^2 
\end{equation}
and $\alpha_0\sim$3 cm$^{-1}$ for the above listed parameters. Thus, for the maximal absorption, when $\hbar\Delta\omega /\varepsilon_l\sim$20 - 30 or $\hbar\Delta\omega\sim |E_0|$, one obtains $\alpha_{max}\sim$45 cm$^{-1}$. Since $\alpha_\omega^{||,\bot}\propto n_{QD}/l^4$, the maximal absorption increases up to $\alpha_{max}\geq 10^3$ cm$^{-1}$ if $n_{QD}>10^{11}$ cm$^{-2}$ and $l\simeq$20 nm; an approximation of low QD concentration remains valid for such a set of parameters. Further increase of $\alpha_{max}$ is possible in the case of heavily doped SL, with a few electrons captured in QD.

Anisotropy of absorption is about 20\% without any strong dependency on effective potential, c. f. Figs. 3a - 3d where parameter $Kl$ varies from 1 to 8. Peculiarities of miniband spectra are visible clearly in $\alpha_\omega^{\bot}$ starting from $Kl\geq$2 while $\alpha_\omega ^{\|}$ does not show any peculiarities at the edges of minibands. This is due to different selection rules for transverse and longitudinally polarized excitations: in the last case, transitions are forbidden at edges of minibands and the spectral dependencies rmain smooth. In addition, Fig. 1a shows a jump of $\alpha_\omega ^{\bot}$ at $\hbar\Delta\omega =0$ which is similar to the jump of the density of states in Fig. 2a (we do not consider IR transitions into shallow underbarrier states at $\hbar\Delta\omega <0$). In Figs. 3b-d, shifts of absorption edges to finite $\hbar\Delta\omega >0$ take place due to lower miniband shifts, see Figs. 1b-d and 2b-d. 

\section{Conclusions}
In summary, we have developed the theory of the superlattice formed by periodically placed quantum dot sheets. It was found that the damping due to random in-plane positions of dots is weak and effect of the sheets on electronic states can be described using of the effective boundary conditions. Within this approach we have demonstrated that the miniband density of states, which describes the interband absorption, and spectra of mid-IR photoexcitation of doped quantum dots into minibands strongly depend on parameters of quantum dot sheets. Visible anisotropy of the absorption coefficient is also found, with transverse absorption which is strongly modulated by the miniband spectrum of SL.

Now we discuss the main assumptions in the calculations performed. We restricted ourselves by the vicinity of $c$-band using the effective-mass approach in Eq. (1) and in further consideration of the photoionization process. In order to describe the energy intervals comparable to the gap, one needs to use the multi-band $\bf kp$-Hamiltonian for more detailed description of QD states. \cite{13} We consider the case of low QD concentration ($n_{QD}/l\sim 10^{15}$ cm$^{-3}$ in our numerical estimates) and the electron-electron interaction effect on the energy spectrum; thus, the IR-absorption should be weak. Numerical estimates for the SL parameters were performed here based on simplified description of QD as an isotropic short-range defect with the binding energy corresponding to typical QD. This approach gives approximate SL parameters only and a more precise description should be based on a numerical solution of the self-consistent Dyson equation taking into account a real potential of QD. \cite {1,13} Because parameters of QD sheet (materials, concentration, and shape of QD) can be very different, such a consideration should be performed for different specific cases (e.g. for Ge/Si-based or A$_{II}$B$_{VI}$-based QD sheets, for review see Ref. 16). 

To conclude, we believe that the results obtained will stimulate an investigation of underbarrier vertical transport of carriers in order to verify SL effect on electronic properties of structures formed by QD sheets. The spectral and polarization dependencies of the mid-IR photoexcitation are convenient for direct measurements because the valence band states are not essential. These results should be important for description of different devices utilizing periodical QD sheet structures.

\section*{ACKNOWLEDGMENT}
This work was supported by the AFOSR. 
%f4
\begin{figure}[tbp]
\begin{center}
\includegraphics{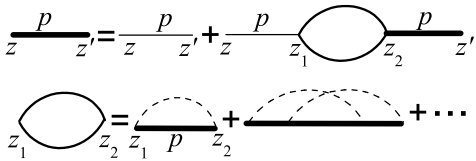}
\end{center}
\addvspace{-0.5 cm}
\caption{Self-consistent Dyson equation for averaged Green's function $G_{Ep}(z,z')$ and the self-energy function $\Sigma_{Ep}\left( z_1 ,z_2\right)$ shown in upper and lower lines, respectively. }
\end{figure}

\appendix*  
\section{}
In order to estimate the corrections beyond the effective potential approach used in Eqs. (9)-(11) we consider here the method of calculations in more details. Using the ${\bf p}z$-representation, one obtains the self-consistent Dyson equation (4) for the averaged Green's function $G_{Ep}(z,z')$ shown by a bold line as it is plotted in Fig. 4. Within the second-order Born approximation, we use the free Green's function in the self-energy function (6) given by the first diagram of the set for $\Sigma_{Ep}$ shown in the lower line of Fig. 4. The next corrections in this set can be neglected under the standard condition \cite{7}
%A1
\begin{equation}
E\gg |\Sigma_{Ep}|\simeq \Lambda  
\end{equation}
and we arrive to Eq. (7) using the free Green's function in $\Sigma_{Ep}$ determined by Eq. (6).
%f5
\begin{figure}[tbp]
\begin{center}
\includegraphics{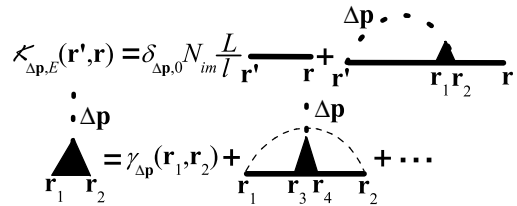}
\end{center}
\addvspace{-0.5 cm}
\caption{Diagram expansion for correlation function ${\cal K}_{\Delta{\bf p}E}({\bf r},{\bf r'})$ written through the diagram set for vertex part shown in lower line.} 
%$\Gamma_{\Delta{\bf p}E}({\bf r},{\bf r'})$. }
\end{figure}

More complicate consideration is necessary for the correlation function $K_{\Delta{\bf p}E}({\bf r},{\bf r'})$ appearing in Eq. (17) because of the random factor $\exp (i\Delta {\bf pR}_{rk}/\hbar )$ describing positions of QDs. Instead Eq. (18) it is convenient to consider the generalized expression
%A2
\begin{equation}
{\cal K}_{\Delta {\bf p},E} ({\bf r}',{\bf r}) = \left\langle {\sum\limits_{rk} {e^{i\Delta {\bf pR}_{rk} /\hbar } {\cal G}_E ({\bf r}',{\bf r})} } \right\rangle 
\end{equation} 
which is shown in Fig. 5. Here a dotted curve corresponds to the averaged factor
%A3
\begin{equation}
\left\langle {\sum\limits_{r_1 k_1 r_2 k_2} {\exp \left( {-\frac{i}{\hbar }\Delta {\bf p} \cdot {\bf R}_{r_1 k_1} } \right)} {\rm  }u\left( {{\bf r}-{\bf R}_{r_2 k_2} } \right)} \right\rangle ,
\end{equation} 
while dashed curves in Figs. 4 and 5 stand for the paired QD potentials. After summation over all reducible diagrams, $K_{\Delta{\bf p}E}({\bf r},{\bf r'})$ is written through the averaged Green's function and the vertex part which is given by the set shown in the lower line of Fig. 5 with the initial vortex determined from (A.3) as follows
%A4
\begin{eqnarray}
\gamma_{\Delta{\bf p}}({\bf r}_1,{\bf r}_2)=n_{QD}\sum_r u\left( -\frac{\Delta {\bf p}}
{\hbar},z_1-rl\right) \\
\times e^{-\frac{i}{\hbar}(\Delta {\bf px}_1+rp_{\bot}l)}\delta ({\bf r}_1-{\bf r}_2) .
\nonumber
\end{eqnarray}
The first correction to Eq. (19) appears, if we use (A.4), as the vertex part in the diagram expansion for correlation function shown in Fig. 5. Performing straightforward calculations under the condition (A.1), one obtains that this correction and next contributions are negligible in comparison with Eq. (19).


\begin{thebibliography}{99}
\bibitem{10}
Using the Koster-Slater approach \cite{8} one can find the binding energy of the short-range defect, $E_0$, from the equation
\begin{equation}
1+U\overline{\rho}_{\varepsilon_a}=\pi U\overline{\rho}_{E_0}/2 \nonumber
\end{equation}
and the normailzed wave function of the ground state $\Psi_P$ used in Eqs. (17) and (20) is given by
\begin{equation}
\Psi_P=\frac{2\sqrt{|E_0 |/2\rho_{|E_0|}}}{|E_0 |+\varepsilon_P} . \nonumber
\end{equation}
\bibitem{12}
According to Ref. 13, the explicite expressions for the reflection coefficient,
$R_{np_\bot}$, and the normalization factor, $\left| \psi_{np_\bot}\right|^2$,
are written as
\begin{equation}
R_{np_\bot}=\frac{\exp (ip_\bot l/\hbar )-\exp (ik_{np_\bot}l)}{\exp (ip_\bot l/\hbar )-\exp (-ik_{np_\bot}l)} \nonumber
\end{equation}
and
\begin{eqnarray}
2l\left| \psi_{np_\bot}\right|^2 =[1-\cos (p_\bot l/\hbar + k_{np_\bot}l)] ~~~~~~~~ \nonumber \\
\times\left\{ 1-\frac{\sin (2k_{np_\bot}l)}{2k_{np_\bot}l} -\cos\frac{p_\bot l}{\hbar} \left[\cos (k_{np_\bot}l)-\frac{\sin (k_{np_\bot}l)}{k_{np_\bot}l} \right]\right\}^{-1} .
\nonumber
\end{eqnarray}
\end{thebibliography}
\end{document}